\documentclass[twoside]{article}

\usepackage{amsthm,amssymb}

\title{Higher spin algebras as higher symmetries}

\author{Xavier Bekaert\thanks{E-mail address: {\tt \footnotesize Xavier.Bekaert@lmpt.univ-tours.fr}}\\
{\small Laboratoire de Math\'ematiques et Physique Th\'eorique}\\
{\small Unit\'e Mixte de Recherche $6083$ du CNRS, F\'ed\'eration Denis Poisson}\\
{\small Universit\'e Fran\c{c}ois Rabelais, Parc de Grandmount}\\
{\small 37200 Tours, France}}

\date{}

\begin{document}

\maketitle

\abstract{The exhaustive study of the rigid symmetries
of arbitrary free field theories is motivated,
along several lines, as a preliminary step
in the completion of the higher-spin interaction problem in full generality. 
Some results for the simplest example (a scalar field) are reviewed 
and commented along these lines.

\vspace{4mm}\noindent\textit{Expanded version of the lectures presented at the ``5th international
school and workshop on QFT \& Hamiltonian systems'' (Calimanesti, May 2006).}}

\vspace{5mm}

\section{Higher-spin interaction problem}

Whereas covariant gauge theories describing arbitrary free
massless fields on constant-curvature spacetimes
of dimension $n$ are firmly
established by means of the unitary representation theory of their
isometry groups, it is still open to question whether non-trivial
consistent self-couplings and/or cross-couplings among those
fields may exist for $n\geqslant 2\,$, such that the deformed gauge algebra
is non-Abelian. The goal of the present paper is to advocate
that a lot of information on the interactions
can be extracted from the symmetries of the free
field theory.

The conventional local free field theories corresponding to unitary
irreducible representations of the helicity group $SO(n-2)\,$, 
that are spanned by completely symmetric tensors, have been constructed a while ago
(for some introductory reviews, see \cite{Intros}). In order to have
Lorentz invariance manifest and second order local field equations with minimal field content,
the theory is expressed in terms of completely symmetric double-traceless tensor gauge
fields $h_{\mu_1\ldots\,\,\mu_s}$
of rank $s>0$, the gauge transformation of which reads
\begin{equation}
\stackrel{(0)}{\delta_\xi}h_{\mu_1\mu_2\ldots\,\mu_s}
=\,\stackrel{(0)}{\nabla}{}_{\mu_1}\,\xi_{\mu_2\ldots\mu_s}\,+\,\mbox{cyclic}\,,
\label{lineartransfos}
\end{equation}
where $\stackrel{(0)}{\nabla}$ is the covariant derivative with
respect to the background Levi--Civita connection and ``cyclic'' stands for the sum of terms necessary to have
symmetry of the right-hand-side under permutations of the indices. The gauge
parameter $\xi$ is a completely symmetric traceless tensor
field of rank $s-1$. In this relativistic field theory, the ``spin" is equal to the
rank $s$. For spin $s=1$ the gauge field $h_\mu$ represents the
photon with $U(1)$ gauge symmetry while for spin $s=2$ the gauge
field $h_{\mu\nu}$ represents the graviton with linearized
diffeomorphism invariance.
The gauge algebra of field independent gauge transformations
such as (\ref{lineartransfos}) is of course Abelian.

Non-Abelian gauge theories for ``lower spin" $s\leqslant 2$ are
well known and essentially correspond to Yang-Mills ($s=1$) and
Einstein ($s=2$) theories for which the underlying geometries
(principal bundles and Riemannian manifolds) were familiar to
mathematicians before the construction of the physical theory. In
contrast, the situation is rather different for ``higher spin"
$s>2$ for which the underlying geometry (if any!) remains obscure.
Due to this lack of information, it is natural
to look for inspiration in the perturbative ``reconstruction"
of Einstein gravity as the non-Abelian gauge theory of
a spin-two particle propagating on a constant-curvature spacetime
(see \textit{e.g.} \cite{Ortin} for a comprehensive review).

Let us denote by $\stackrel{(0)}{S}[\,h_{\mu_1\ldots\,\mu_s}]$
the Poincar\'e-invariant, local, second-order, quadratic,
ghost-free, gauge-invariant action of a spin-$s$ symmetric tensor
gauge field. In order to perform a perturbative analysis via the
Noether method \cite{Noether}, the non-Abelian interaction problem for a
collection of higher
(and possibly lower) spin gauge fields is formulated as a
deformation problem.

\vspace{1mm}\noindent{\bfseries{Higher-spin interaction problem:}}
\textit{List all Poincar\'e-invariant local deformations
$$S[h]\,=\,\stackrel{(0)}{S}[h]\,+\,\varepsilon\,\stackrel{(1)}{S}[h]
\,+\,{\cal O}(\varepsilon^2)$$
of a positive sum, with at least one $s>2$,
$$\stackrel{(0)}{S}[h]\,=\,\sum_{s}\stackrel{(0)}{S}[\,h_{\mu_1\ldots\,\mu_s}]$$
of quadratic actions such that the deformed local gauge symmetries
$$\delta_\xi h\, =\,\stackrel{(0)}{\delta_\xi}h \,+\,\varepsilon\,\stackrel{(1)}{\delta_\xi}h\,
+\,{\cal O}(\varepsilon^2)$$
are already non-Abelian at first order, in the deformation parameters $\varepsilon$ and do not arise from local
redefinitions
$$h\, \rightarrow\,h \,+\,\varepsilon\,\phi(h)\,+\,{\cal O}(\varepsilon^2)\,,\quad
\xi\, \rightarrow\,\xi
\,+\,\varepsilon\,\zeta(h,\xi)\,+\,{\cal O}(\varepsilon^2)$$ of the
gauge fields and parameters.}\vspace{1mm}

This well-posed mathematical problem is expected to possess non-trivial solutions including higher-spin fields,
as strongly indicated by Vasiliev's works (for some reviews, see \cite{Vasiliev} and references therein)
and deserves to be investigated further along systematic lines.

\section{The Noether method}

The assumption that the deformations are formal power series in
some deformation parameters $\varepsilon$ enables to investigate the problem
order by order. The crucial observation of any perturbation theory
is that the first order deformations are constrained by the
symmetries of the undeformed system. In the present case, the
Noether method scrutinizes the gauge symmetry of the action,
$\delta_\xi S=0\,$. At zeroth order, the latter equation is
satisfied by hypothesis. At first order, it reads
\begin{equation}
\stackrel{(0)}{\delta_\xi}\stackrel{(1)}{S_{}}\,
+\,\stackrel{(1)}{\delta_\xi}\stackrel{(0)}{S_{}}\,=\,0\,.
\label{firstorder}
\end{equation}
This equation may be used to constrain the possible deformations
by reinterpreting them as familiar objects of the undeformed gauge
theory.

By definition, an \textit{observable} of a gauge theory is
a functional which is gauge-invariant on-shell, while
a \textit{reducibility parameter} of a gauge theory is a
gauge parameter such that the corresponding gauge variation
vanishes off-shell.

\vspace{1mm}
\noindent{\bfseries{First-order deformations in terms
of the undeformed theory:}}

\noindent $\bullet$ \textit{First-order deformations of the action
are observables of the undeformed theory.}

\noindent $\bullet$ \textit{First-order deformations of the gauge
symmetries evaluated at reducibility parameters of the undeformed gauge
theory define symmetries of the undeformed theory.}

\noindent{\bfseries{Proof:}} In (\ref{firstorder}) the infinitesimal variation
$\,\stackrel{(1)}{\delta_\xi}\stackrel{(0)}{S_{}}$ of the
undeformed action is proportional to the undeformed
Euler--Lagrange equations. This proves the fist part of the theorem.
Reducibility parameters
$\overline{\xi}$ of the undeformed gauge theory verify $\stackrel{(0)}{\delta_{\overline{\xi}}}h=0$
by definition.
Inserting this fact into (\ref{firstorder}) with
$\xi=\overline{\xi}$ gives
$\stackrel{(1)}{\delta_{\overline{\xi}}}\stackrel{(0)}{S_{}}=0\,$,
which is precisely the translation of the second part of the
theorem. \qed

\vspace{1mm} In the mathematical litterature,
a \textit{(conformal) Killing tensor} of a pseudo-Riemannian manifold
is a symmetric tensor field $\xi$ such that its symmetrized
covariant derivative with respect to the Levi--Civita
connection,
$\nabla_{\mu_1}\,\xi_{\mu_2\ldots\mu_s}\,+$ cyclic,
vanishes (modulo a term proportional to the metric
for conformal Killing tensors).
Therefore, any reducibility parameter $\overline{\xi}$ of the spin-$s$ symmetric gauge
field theory on the constant-curvature spacetime $\cal M$
is identified with a Killing tensor of rank $s-1$
of the manifold $\cal M$.
The space of Killing tensors on any constant-curvature spacetime
is known to be finite-dimensional
\cite{Killingconstcurvv}, thus the linear gauge symmetries
(\ref{lineartransfos}) are irreducible.

These results suggest two strategies for
addressing the higher-spin interaction problem. The most ambitious one
is the computation of \textit{all} local observables of the free
gauge theory associated to deformations of the gauge algebra. This
result would provide the exhaustive list of algebra-deforming first order vertices,
but this computation is technically demanding and seems out of
reach in the completely general case. Nevertheless, the BRST
reformulation of the problem \cite{BH} allowed the complete
classification of non-Abelian deformations in various particular
cases (see \textit{e.g.} the review \cite{BBCL} and references therein).
Actually, a more humble strategy is the computation of \textit{all}
rigid symmetries of the free irreducible gauge theory.
It is of interest because the knowledge of these rigid symmetries would
strongly constrain the candidates for gauge symmetry deformations.
Indeed, the constant tensors appearing in the rigid symmetries
could be compared with the complete list \cite{Killingconstcurvv}
of constant-curvature spacetime Killing tensors.

\section{Free theory symmetries}

Bosonic fields are usually described in terms of their components
living in some subspace $V$ of the space $\otimes({\mathbb R}^n)$
of tensors on ${\mathbb R}^n$ (\textit{e.g.}
$V=\odot({\mathbb R}^n)$ for symmetric tensor fields). The background metric of the
constant-curvature spacetime induces some non-degenerate bilinear
form on $V$. This defines a non-degenerate sesquilinear form
$\langle\,\,\,\mid\,\,\,\rangle$ on the space $L^2({\mathbb R
}^n)\otimes V$ of square-integrable fields taking values in the
countable space $V$ (the components). Let $\dagger$ stands for the
adjoint with respect to the sesquilinear form $\langle\,\,\,\mid\,\,\,\rangle\,$.

Any quadratic action for
bosonic fields $\psi$ can be expressed as a quadratic form
\begin{equation}
\stackrel{(0)}{S_{}}[\psi]\,=\,\frac12\,\langle\,\psi\mid\textsc{K}\mid\psi\,\rangle\,,
\label{quadract}
\end{equation}
where the kinetic operator $\textsc{K}$ is self-adjoint,
$\textsc{K}^\dagger=\textsc{K}$. Because the sesquilinear form
$\langle\,\,\,\mid\,\,\,\rangle$ is non-degenerate, the
Euler-Lagrange equation extremizing the quadratic action is the
linear equation
\begin{equation}
\frac{\delta
\stackrel{(0)}{S_{}}}{\delta\langle\,\psi\mid}\,=\,\textsc{K}|\,\psi\,\rangle\,=\,0\,.
\label{ELequ}
\end{equation}
Moreover, the quadratic form
$\langle\,\psi\mid\textsc{K}\mid\psi\,\rangle$ is degenerate if
and only if the kinetic operator $\textsc{K}$ is degenerate. This
happens if and only if there exists a linear operator $\textsc{P}$
(on $L^2({\mathbb R}^n)\otimes V$) such
that $\textsc{K}\,\textsc{P}=0$. Infinitesimal gauge symmetries
then read
$$\stackrel{(0)}{\delta_\chi}\,\mid\psi\,\rangle
\,=\,\textsc{P}\mid\chi\,\rangle\,,$$ with gauge parameters
$\chi\,$. The Noether identity is
$\textsc{P}^\dagger\textsc{K}=(\textsc{K}\,\textsc{P})^\dagger=0\,$.

A \textit{symmetry of the quadratic action}
(\ref{quadract}) is an invertible linear pseudo-differential operator $\textsc{U}$ preserving the
quadratic form
$\langle\,\,\mid\textsc{K}\mid\,\,\rangle$. In other words,
$$\textsc{U}^\dagger\,\textsc{K}\,\textsc{U}\,=\,\textsc{K}\,.$$
The \textit{group of off-shell symmetries} is the group of
symmetries of the quadratic action endowed with the composition
$\circ$ as product. A \textit{symmetry generator of the quadratic
action} (\ref{quadract}) is a linear differential operator $\textsc{T}$ which
is self-adjoint with respect to the quadratic form
$\langle\,\,\mid\textsc{K}\mid\,\,\rangle$. More concretely,
$$\textsc{K}\,\textsc{T}\,=\,\textsc{T}^\dagger\textsc{K}\,.$$
Any symmetry generator $\textsc{T}$ defines a symmetry
$\textsc{U}=e^{i\textsc{T}}$ of the quadratic action
(\ref{quadract}). If $\textsc{T}=\textsc{T}^\dagger$ then the
linear operator $\textsc{T}$ is a symmetry generator of the
quadratic action if and only it commutes with $\textsc{K}$. The
\textit{real Lie algebra of off-shell symmetries} is the algebra
of symmetry generators of the quadratic action endowed with $i$
times the commutator as Lie bracket,
$\{\,\,,\,\,\}\,:=\,i\,[\,\,,\,\,]$.

A \textit{symmetry of the linear equation} (\ref{ELequ})
is a linear differential operator $\textsc{T}$ obeying
\begin{equation}
\textsc{K}\,\textsc{T}\,=\,\textsc{S}\,\textsc{K}\,,
\label{symgen}
\end{equation}
for some linear operator $\textsc{S}$. Such a symmetry
$\textsc{T}$ preserves the space Ker$\textsc{K}$ of solutions to
the equations of motion. Any symmetry generator $\textsc{T}$ of
the action (\ref{quadract}) is always a symmetry of the equation
of motion (\ref{ELequ}) with $\textsc{S}=\textsc{T}^\dagger$ in
(\ref{symgen}). A symmetry $\textsc{T}$ is \textit{trivial
on-shell} if $\textsc{T}=\textsc{R}\textsc{K}$ for
some linear operator $\textsc{R}$. Such an on-shell-trivial
symmetry is always a symmetry of the field equation (\ref{ELequ}),
since it obeys (\ref{symgen}) with
$\textsc{S}=\textsc{K}\textsc{R}$.
The algebra of on-shell-trivial symmetries obviously forms a left
ideal in the algebra of linear differential operators endowed with the
composition $\circ$ as multiplication. Furthermore, it is also a right
ideal in the algebra of symmetries of the linear equation
(\ref{ELequ}). The \textit{complex associative algebra of on-shell
symmetries} is the associative algebra of symmetries of the linear
equation quotiented by the two-sided ideal of on-shell-trivial
symmetries. The \textit{complex Lie algebra of on-shell
symmetries} is the algebra of on-shell symmetries
endowed with the commutator as Lie bracket.

Notice that when $\textsc{K}$ is non-degenerate,
a linear operator $\textsc{T}=\textsc{R}\textsc{K}$
is a symmetry generator of the quadratic action
(\ref{quadract}) if and only if $\textsc{R}$ is self-adjoint.
Moreover, the Lie subalgebra of such on-shell-trivial
symmetry generators is an ideal in the Lie algebra of
off-shell symmetries.

\section{Higher-spin algebras}

Let $\mathfrak{g}$ be the Lie algebra corresponding to the
finite-dimensional
(conformal) isometry group $G$ of the constant-curvature spacetime of
dimension $n>2$. For $n=2\,$, the spacetime may be arbitrary
and the conformal algebra is of course infinite-dimensional.
If the free field theory is relativistic, then $\mathfrak{g}$
is linearly realized on the space $L^2({\mathbb R}^n)\otimes V$
(respectively, Ker$\textsc{K}$) of off-shell (resp. on-shell)
fields. This induces a linear realization of the universal
enveloping algebra ${\cal U}(\mathfrak{g})$ over $\mathbb C$. The real form of this realization 
corresponding to the self-adjoint
operators, endowed with $i$ times the
commutator as Lie bracket, is nowadays referred to
as \textit{(conformal) on/off-shell higher-spin algebra of the
constant-curvature spacetime} (see e.g. \cite{X} for an elementary introduction to
such algebraic structures).
The name comes from the fact that its generators are in
``higher-spin" representations of the Lorentz group,
and the algebra is said to be ``on" or ``off" shell whether
the algebra is realized on the space of solutions of
the Euler-Lagrange equations or not.

The isometry algebra $\mathfrak{g}$ of a constant-curvature
spacetime is a module of the Lorentz subalgebra $\mathfrak{o}(n-1,1)\subset\mathfrak{g}$
for the adjoint representation.
This module decomposes as the sum of
two irreducible $\mathfrak{o}(n-1,1)$-modules:
the ``translations" are in the vector module $\cong {\mathbb R}^n$ while
the boosts and rotations are in the antisymmetric module $\cong \wedge^2({\mathbb R}^n)$.
These representations are labelled by one-column Young diagrams of,
respectively, one and two cells. The number of columns is associated with the spin.
The fact that the generators of ${\cal U}(\mathfrak{g})$ are in
higher-spin representations is summarized in the following result.

\vspace{1mm}\noindent{\bfseries{Universal enveloping algebra of
isometries:}} \textit{The universal enveloping algebra ${\cal
U}(\mathfrak{g})$ of the isometry algebra $\mathfrak{g}$ of an
$n$-dimensional constant-curvature spacetime is an
infinite-dimensional module of the general linear Lie algebra
$\mathfrak{gl}(n)$, decomposing as an infinite sum of
finite-dimensional irreducible $\mathfrak{gl}(n)$-modules
labelled by the set of all Young diagrams, with multiplicity one,
the first column of which has length $\leqslant n$.}

\vspace{1mm}\noindent{\bfseries{Proof:}} The
Poincar\'e-Birkhoff-Witt theorem states that the universal
enveloping algebra ${\cal U}(\mathfrak{g})$ is isomorphic to the symmetric
algebra $\odot(\mathfrak{g})$ as a vector space. As a
$\mathfrak{gl}(n)$-module, the vector space $\mathfrak{g}$
is isomorphic to
the sum ${\mathbb R}^n\oplus\wedge^2({\mathbb R}^n)$ of
irreducible modules.
This leads to the following isomorphism of
modules:
\begin{equation}
\odot(\mathfrak{g})\,\cong\,\Big(\odot({\mathbb R}^n)\Big)
\otimes\Big(\odot\big(\wedge^2({\mathbb R}^n)\big)\Big)\,.
\label{Kronecker}
\end{equation}
The idea is to evaluate the right-hand-side of (\ref{Kronecker})
using the available technology on Kronecker
products of irreducible representations \cite{Littlewood}.
The module $\odot({\mathbb R}^n)$ decomposes as the
infinite sum of irreducible modules labelled by all one-row Young diagrams
with multiplicity one.
A formula of Littlewood for symmetric plethsyms implies that the
module $\odot\big(\wedge^2({\mathbb R}^n)\big)$
decomposes as the infinite sum of irreducible modules,
with multiplicity one, labelled by all Young diagrams
with columns of even lengths.
The Kronecker product in (\ref{Kronecker}) decomposes as the infinite sum
of all the Kronecker products
between a one-row Young diagram and
a Young diagram with columns of even lengths, each with multiplicity one.
Using the Littlewood--Richardson rule, one may show that the result
of this computation is the infinite sum of irreducible modules
labelled with all possible Young diagrams, each with multiplicity one.
The Young diagrams whose first column has length greater than $n$
lead to vanishing modules, hence they do not appear in the series.
\qed

\vspace{2mm}The higher-spin algebras are important in relativistic field theories
because they always appear as spacetime symmetry algebras
in the free limit.

\vspace{1mm}\noindent{\bfseries{Spacetime symmetries of
relativistic free field theories:}} \textit{If the Lie algebra of
off/on-shell symmetries contains the (conformal) isometry algebra
$\mathfrak{g}$ of some constant-curvature spacetime $\cal M$, then
it also contains the (conformal) off/on-shell higher-spin algebra
of $\cal M$.}

\vspace{1mm}\noindent{\bfseries{Proof:}}
The Poincar\'e-Birkhoff-Witt theorem states that one can realize
the universal enveloping ${\cal U}(\mathfrak{g})$
as Weyl-ordered polynomials in the elements of
the Lie algebra $\mathfrak{g}$.
The above theorem is proved by observing that any Weyl-ordered polynomial
in on-shell symmetries is itself an on-shell symmetry. As observed in \cite{Mikhailov}, the same is
true for symmetry generators.\qed

As an important corollary, the theorem implies that any relativistic
\textit{free} field theory has an infinite number of rigid symmetries, and
therefore it possesses an infinite number of conserved currents via the
Noether theorem, as it is well known. 
Notice that relativistic \textit{integrable}
models are precisely such that they possess an infinite set of commuting rigid symmetries
corresponding to an infinite set of conserved charges in involution.
The infinite-dimensional subalgebra of symmetries of the free field theory generated by the translations only is, of course, Abelian. 
Actually, the factorization property is deeply related to the preservation of this subalgebra of symmetries at the interacting level \cite{Zamolodchikov}.
Thus the relationship between higher-spin algebras
and integrable models appears to be very intimate
(see also \cite{Vasiliev:1995} and references therein). The strong form of the Maldacena conjecture (in the large $N$ limit)
and the integrability properties recently enlightened in this context are further indications of such a relationship.

Symmetries may be characterized by their action
on the spacetime coordinates. A smooth change of coordinates is
generated by a first-order linear differential operator.
Therefore, a higher-order linear differential operator does not
generate coordinate transformations.
For instance, an isometry generator is a first-order
linear differential operator corresponding to a Killing vector field,
but the spacetime higher-symmetries are powers of such isometry generators,
hence they are higher-order linear differential operators.
They do not generate coordinate transformations
and this explains why spacetime higher-symmetries
are usually not considered in textbooks.

Let us focus on the first non-trivial example
of free field theory: the quadratic action of
a complex scalar field on an $n$-dimensional spacetime
$\cal M$. In such case, the space $V=\mathbb C$ and the
kinetic operator $\textsc{K}$ can be taken to be
a constant mass term plus the Laplacian on $\cal M$,
$$
\stackrel{(0)}{\Box}\,=\,\stackrel{(0)}{\nabla}{}^{\mu}\stackrel{(0)}{\nabla}_{\mu}\,.
$$
A scalar field is said to be \textit{conformal} if its kinetic operator is the conformal
Laplacian
\begin{equation}
\stackrel{(0)}{\Box}\,-\,\frac{n-2}{4\,(n-1)}\,\stackrel{(0)}{R}\,,
\end{equation}
where $R$ denotes the scalar curvature.
The quadratic action and the linear equation are symmetric
under the full conformal algebra $\mathfrak{o}(n,2)$
if and only if the scalar field is conformal and has conformal weight $1-n/2$.

\vspace{1mm}\noindent{\bfseries{Higher symmetries of the conformal scalar field:}}
\textit{For the quadratic action of a complex
conformal scalar field on a constant-curvature spacetime
$\cal M$ of dimension $n\geqslant 2$, the following spaces over $\mathbb R$ are isomorphic:}

\noindent $\bullet$ \textit{The Lie algebra of off-shell
symmetries quotiented by the ideal of on-shell-trivial symmetry generators,}

\noindent $\bullet$ \textit{A real form of the associative algebra of on-shell symmetries.}

\noindent $\bullet$ \textit{The conformal on-shell higher-spin algebra,}

\noindent $\bullet$ \textit{The real algebra of Weyl-ordered polynomials
in the conformal Killing vector fields
quotiented by the ideal generated by the conformal Laplacian,
endowed with $i$ times the commutator as Lie bracket.
The symbols of these differential operators, $$\textsc{T}=(-i)^r\overline{\xi}^{\mu_1\ldots\mu_r}(x)\stackrel{(0)}{\nabla}_{\mu_1}\ldots
\stackrel{(0)}{\nabla}_{\mu_r}\,+\,\mbox{lower}\,+\,\mbox{on-shell-trivial}\,,$$
may be represented by real traceless symmetric tensor fields $\overline{\xi}$
which are conformal Killing tensors.}

\noindent\textit{Moreover, in $n=2$ dimensions
the theorem is valid for an arbitrary spacetime manifold.}

\vspace{1mm}\noindent{\bfseries{Proof:}}
The theorem can be extracted from the results of \cite{Eastwood}
on flat spacetime of dimension $n>2$ by taking into account that any constant-curvature
spacetime $\cal M$ can be seen as a
conic in the projective null cone of the ambient space ${\mathbb R}^{n,2}\,$.
The two-dimensional case is addressed by using the left/right-moving coordinates.
\qed

\vspace{2mm} Notice that the on-shell higher-spin algebra of a non-conformal scalar field 
on a constant-curvature spacetime is a proper
subalgebra of the universal enveloping algebra of the isometry algebra
$\mathfrak{g}$:
it decomposes as the infinite sum of irreducible
$\mathfrak{o}(n-1,1)$-modules labelled by all two-row Young
diagrams with multiplicity one, as reviewed in \cite{Vasiliev,BBCL}.
This algebra is in one-to-one
correspondence with the space of reducibility parameters of the
infinite tower of symmetric tensor gauge fields where each field
appears once and only once for each given spin $s>0$. Moreover,
notice that the $AdS_{n+1}/CFT_n$ correspondence for $n>2$ in the weak
tension/coupling limit also makes use of the isomorphism between
the on-shell higher-spin algebra of a non-conformal scalar field on
$AdS_{n+1}$ and the on-shell symmetry algebra of a conformal scalar
field on ${\mathbb R}^{n-1,1}$ (see \cite{Konstein} for the correspondence
at the level of conserved currents).
Remark also that the conformal on-shell higher-spin algebra
of a two-dimensional spacetime for a massless scalar field
is isomorphic to the direct sum of $\mathfrak{u}(1)$ and the
two Lie algebras of differential operators
for the left and right moving sectors respectively.
Each of such algebras of differential operators
is isomorphic to the algebra ${\cal W}_\infty$
with zero central charge \cite{Bakas}.

\vspace{2mm} The deep connection between higher-spin algebras and integrable models is exhibited
by the following example in $n=2$ dimensions.

\vspace{1mm}\noindent{\bfseries{Higher symmetries of the interacting scalar field:}} 
\textit{A non-linear action of a real
scalar field on the two-dimensional Minkowski spacetime,
without derivative interaction term, of the form}
$$S[\phi]=\,\frac12\,\langle\,\phi\mid\Box\mid\phi\,\rangle + \int d^2x \,V(\phi)\,,\quad\quad
V(\phi)={\cal O}(\phi^2)\,,$$
\textit{is invariant under an infinite number of local infinitesimal rigid symmetry transformations,
independent of the coordinate $x^\mu$, if and only if $$V(\phi)=\pm\left(\frac{m}{\alpha}\right)^2\big(\cos\,(\alpha\, \phi)-1\big)\,,\quad\quad m\in{\mathbb R}\,,$$ the parameter $\alpha$ is either purely real or imaginary. 
In such case, the field $\phi$ either corresponds to a free massless
scalar field ($m=0$), a free massive scalar field ($m\neq 0\,$, $\alpha= 0$) or 
sine-Gordon theory ($m\neq 0\,$, $\alpha\neq 0$).}

\textit{Moreover, via linearisation, there is a one-to-one correspondence between:}

\noindent $\bullet$ \textit{The set of on-shell non-trivial, polynomial in the field derivatives, coordinate-independent, symmetry transformations 
of the sine-Gordon Lagrangian}

\noindent $\bullet$ \textit{The Lie algebra of coordinate-independent off-shell
symmetries of a free real scalar field quotiented by the ideal of on-shell-trivial symmetry generators,}

\noindent $\bullet$ \textit{A proper Abelian Lie subalgebra of the on-shell higher-spin algebra
of the Minkowski plane,}

\noindent $\bullet$ \textit{The space of harmonic odd polynomials
in the momenta $\textsc{P}_\mu=-i\partial_\mu\,$. These differential operators $\textsc{T}$
may be represented by real traceless symmetric constant tensors
$\lambda$:}
$$\textsc{T}=i\,\lambda^{\mu_1\ldots\mu_{2q+1}}{\partial}_{\mu_1}\ldots{\partial}_{\mu_{2q+1}}\,
+\,\mbox{on-shell-trivial}\,.$$

\vspace{1mm}\noindent{\bfseries{Proof:}} The first part of the theorem is a straightforward consequence of the 
results of \cite{Dodd} in the case when $V(\phi)$ is at least quadratic in $\phi$ (by hypothesis).
The second part is proven by selecting all coordinate-independent symmetries 
of a free real scalar field and comparing them with the conserved currents of \cite{Dodd}.
In both cases, the Noether correspondence between non-trivial conserved currents and non-trivial symmetries 
(see \textit{e.g.} \cite{Barnich} for a precise statement of this isomorphism)
is performed via the Hamiltonian formulation of a two-dimensional scalar field where one of the light-cone coordinate 
plays the role of ``time.''
\qed

\section{A gauge principle for higher-spins ?}

The analogy with lower-spins suggests to guess the full
non-Abelian gauge theory by making use of the ``gauge principle."
Moreover, this point of view actually provides
a concrete motivation for using the higher-spin algebras
in the interaction problem.

The idea is to consider some ``matter" system described by a quadratic
action (\ref{quadract}) with some algebra of rigid symmetries.
The rigid symmetries $\textsc{U}$ of this system are by definition
in the ``fundamental" representation of the algebra of off-shell symmetries
of the action (\ref{quadract}).
Connections are usually introduced in order to ``gauge" these rigid
symmetries by allowing $\textsc{U}$ to be a smooth function
on ${\mathbb R}^n$ taking values in the group of off-shell symmetries
of the action (\ref{quadract}).
In order to construct a covariant derivative $D=\partial+\Gamma$, one
introduces a connection defined as a covariant vector field
$\Gamma_\mu$ taking values
in the Lie algebra of off-shell symmetries and transforming as
\begin{equation}
\mid\psi\,\rangle\longrightarrow \textsc{U}\mid\psi\,\rangle\,,\quad
\Gamma\longrightarrow \textsc{U}\,D\,\textsc{U}^{-1}\,,
\label{gauges}
\end{equation}
in such a way that
$$D\mid\psi\,\rangle\longrightarrow \textsc{U}\,D\mid\psi\,\rangle\,.$$
The minimal coupling is the replacement
of all partial derivatives $\partial$ in the kinetic operator $\textsc{K}(\partial)$
by covariant derivatives $D$ which should ensure that the quadratic action
$\langle\,\psi\mid\textsc{K}(D)\mid\psi\,\rangle$
is preserved by gauge symmetries (\ref{gauges}).
The connection transforms in the ``adjoint'' representation
of the rigid symmetries while the matter field transforms in
the ``fundamental.'' (More precisely, the covariant derivative transforms in the adjoint
while the matter field belongs to a module of the gauge algebra.)

The introduction of a connection requires the introduction
of some new dynamical fields: the ``gauge" sector.
In Yang-Mills gauge theories, the rigid symmetry is internal
and the connection is itself made of spin-$1$ gauge fields.
For spacetime symmetries, the relation between the connection
and the gauge field is more complicated. For instance,
in Einstein gravity the Levi-Civita connection
is expressed in terms of the first derivative of the metric
via the torsionlessness and metricity constraints.
In general, the spin-$s$ tensor field propagating
on a constant-curvature spacetime
is expected to be the perturbation of some background field
$$g_{\mu_1\ldots\mu_s}\,=\,\stackrel{(0)}{g}_{\mu_1\ldots\mu_s}
\,+\,\varepsilon\,h_{\mu_1\ldots\mu_s}\,,$$
so that the deformed gauge symmetries would be of the form
\begin{equation}
\delta_\xi g_{\mu_1\mu_2\ldots\,\mu_s}
=\,\varepsilon\,(D\xi)_{\mu_1\mu_2\ldots\,\mu_s}\,,
\label{full}
\end{equation}
where the covariant derivative $D=\nabla+{\cal O}(\varepsilon)$ starts as
the covariant derivative with respect to the Levi--Civita
connection for the spacetime metric plus non-minimal corrections.
Thus the background connection is identified with the Levi-Civita connection
for the background metric, and the linearization of (\ref{full})
reproduces (\ref{lineartransfos}).
Furthermore, the reducibility parameters of (\ref{lineartransfos})
exactly correspond to the gauge symmetries
(\ref{full}) leaving the background geometry invariant.
In the present case, this group of rigid symmetries contains
the isometry group $\mathfrak{g}$ of the constant-curvature spacetime.
The classical theory of (in)homogeneous pseudo-orthogonal groups
tells us that completely symmetric tensor fields which are invariant
under $\mathfrak{g}$ are constructed from products of the background metric: $$\stackrel{(0)}{g}_{(\mu_1\mu_2}\ldots\stackrel{(0)}{g}_{\mu_{2m-1}\mu_{2m})}\,.$$
Thus, along these lines, only even-spin
symmetric tensor fields can be perturbations of
a non-vanishing higher-spin background in a constant-curvature spacetime.
The first-order deformation of the gauge symmetries (\ref{lineartransfos})
following from (\ref{full}) would be of the schematic form
\begin{equation}
\stackrel{(1)}{\delta_\xi}h_{\mu_1\mu_2\ldots\,\mu_s}
=\,(\,\stackrel{(1)}{\Gamma}\cdot\,\,\xi\,)_{\mu_1\mu_2\ldots\,\mu_s}\,,
\label{firstorderdef}
\end{equation}
where $\stackrel{(1)}{\Gamma}$ stands for the linearized
connection (including the linearized Levi-Civita connection) and
the dot stands for the action on the gauge parameter $\xi$. 
The transformations (\ref{firstorderdef}) evaluated
on Killing tensors $\overline{\xi}$ of the background spacetime
would be rigid symmetry transformations of the free gauge theory. 
This property highly constrains the possible expressions
for the linearized connection.

Let us now consider the expansion of the minimally coupled action for 
the ``matter'' sector in power series of $\varepsilon\,$:
$$\frac12\langle\,\psi\mid\textsc{K}(D)\mid\psi\,\rangle\,=\,\frac12\langle\,\psi\mid\textsc{K}(\partial)\mid\psi\,\rangle\,+\,
\varepsilon\,\langle\,h\mid J\,\rangle\,+\,{\cal O}(\varepsilon^2)\,,$$
where $J$ denotes a set of symmetric tensors which are bilinear in $\psi$ and their derivatives.
Assuming that the ``matter'' sector is strictly distinct from the ``gauge'' sector, the gauge invariance of the complete action at first order in $\varepsilon$ requires 
the symmetric tensors $J^{\mu_1\mu_2\ldots\,\mu_s}$ to be conserved up to terms
proportional to the ``matter'' free field equations (and derivatives thereof)
corresponding to first-order deformations
\begin{equation}
\stackrel{(1)}{\delta_\xi}\mid\psi\,\rangle\,=\,\stackrel{(1)}{\textsc{U}}\mid\psi\,\rangle\,
\label{lingauges}
\end{equation}
of the gauge transformations of the ``matter'' sector, where $\stackrel{(1)}{\textsc{U}}$ is a linear differential operator depending linearly on $\xi$ and its derivatives. At zeroth order in $\epsilon\,$, the ``gauge'' group does not act on the matter.
Therefore, at leading order, the transformation law (\ref{gauges}) reads as (\ref{lingauges}). Via the Noether correspondence, the space of all rigid symmetries of the ``matter'' quadratic action determines the space of all on-shell-conserved currents bilinear in the ``matter'' fields. The latter ones determine, at first order, the ``fundamental'' representation of the ``gauge'' group. The transformations (\ref{lingauges}) evaluated on Killing tensors $\overline{\xi}$ must define off-shell symmetries of the ``matter'' quadratic action. Their algebra algebra is non-Abelian, hence the ``gauge'' algebra is already non-Abelian at first order.

As a suggestive example, one may consider a ``matter'' sector containing only a single scalar field.

\vspace{1mm}\noindent{\bfseries{Noether cubic couplings of a scalar field:}}
\textit{The minimally coupled action of a complex
scalar field on flat spacetime, given by}
$$S[\phi]=\,\frac12\,\langle\,\phi\mid\Box- m^2\mid\phi\,\rangle -\,\varepsilon\,\int d^nx\,\,h_{\mu_1\mu_2\ldots\,\mu_s}J^{\mu_1\mu_2\ldots\,\mu_s}
+\,{\cal O}(\varepsilon^2)\,,$$
\textit{is invariant at first order in $\varepsilon\,$, for any symmetric tensor field $\xi^{\mu_1\mu_2\ldots\,\mu_{s-1}}\,$, 
under infinitesimal symmetry transformations 
$$
\delta_\xi h_{\mu_1\mu_2\ldots\,\mu_s}
=\,\,\stackrel{(0)}{\delta_\xi}h_{\mu_1\mu_2\ldots\,\mu_s}\,+\,{\cal O}(\varepsilon)\,,
$$
and
\begin{equation}
\delta_\xi\mid\phi\,\rangle\,=\,\varepsilon\,\textsc{T}\mid\phi\,\rangle+\,{\cal O}(\varepsilon^2)\,,
\label{diffop}
\end{equation}
where the symbol of the differential operator $\textsc{T}$ is represented by $\xi$ and the lower order terms depend on derivatives of $\xi$,
$$\textsc{T}=(-i)^{s-1}\xi^{\mu_1\ldots\mu_{s-1}}\partial_{\mu_1}\ldots
\partial_{\mu_{s-1}}\,+\,\,\mbox{lower}+\mbox{on-shell-trivial}\,,$$
if and only if the on-shell-conserved current $J$
is equivalent to a Noether current associated to the coordinate-independent off-shell symmetries of the free scalar field. 
This defines a one-to-one correspondence between equivalence classes of such symmetric Noether currents $J\,$, bilinear in $\phi$ and its derivatives, and equivalence classes of such deformations $\delta_\xi\phi$ at first order.}

\vspace{1mm}\noindent{\bfseries{Proof:}}
The explicit equation expressing the gauge invariance of the minimaly coupled action for any symmetric tensor field $\xi(x)$ of rank $s-1$ 
precisely states that the symmetric tensor $J$ of rank $s$ is conserved modulo terms proportional to field equation of the scalar field $\phi$.
The one-to-one correspondence, precisely explained in \cite{Barnich}, between equivalence classes of on-shell conserved currents and equivalence classes of off-shell symmetry transformations shows explicitly that $J$ is necessarily related to a coordinate-independent transformation of the form (\ref{diffop}).
In turn, these transformations are obtained by evaluating the transformation (\ref{diffop}), at lowest order in $\varepsilon$ and on gauge parameters $\xi$ equal to constant Killing tensors.
The sufficiency is proven by making use of the symmetric conserved currents of \cite{current}. The second part of the theorem follows from the fact that trivial currents define trivial deformations and conversely, as it can be seen explicitly.  
\qed

\vspace{2mm}In the lower-spin case, one recovers the standard minimal coupling procedure. For $s=1\,$, the minimal coupling stops at second order in $\varepsilon$
since $J^\mu$ is the $U(1)$ current and $h_\mu$ is the Abelian vector gauge field. For $s=2\,$, the minimal coupling at first order is the usual coupling between a spin-two gauge field and the energy-momentum tensor $J^{\mu\nu}$ leading to the coordinate transformations of the scalar field, generated by the vector fields $\textsc{T}=-i\,\xi^\mu(x)\,\partial_\mu\,$. The commutators of such infinitesimal transformations close and define the Lie bracket of vector fields, so the underlying gauge symmetry algebra may already be guessed at first order for gravity: it is the Lie agebra of smooth vector fields, \textit{i.e.} the Lie algebra for the group of diffeomorphisms.
The minimally coupled action is obtained to all orders by introducing the Levi-Civita connection.

In the higher-spin case, it should be stressed that the trace conditions on the gauge field and parameter have not been stated in the former proposition because they may indeed be relaxed in order to simplify its formulation. (Nevertheless, these constraints may be included by consistently imposing weaker conservation laws on double-traceless currents.)
Moreover, it is convenient to remove trace constraints for searching a Non-Abelian higher-spin gauge symmetry algebra. Actually, the trace constraints may be removed for free field theories in several ways (see \cite{FS} for some reviews, and \cite{FMS} for the latest developments). The Lie algebra of gauge transformations (\ref{diffop}) for the infinite tower of all gauge parameters ($1\leqslant s<\infty$) is a real form of the algebra of linear differential operators 
on ${\mathbb R}^n$ endowed with $i$ times the commutator as Lie bracket. Notice also that the unital associative algebra of linear differential operators on ${\mathbb R}^n$ is isomorphic to the universal enveloping algebra of vector fields on ${\mathbb R}^n\,$.
(Strictly speaking, this is true only for \textit{polynomial} vector fields and differential operators, 
more sophisticated mathematical statements may be required for \textit{smooth} functions, 
but this point is only technical.)
More concretely, the symbol of a differential operator of order $r$ is represented by a symmetric tensor field of rank $r$.
In the light of these remarks, it is tempting to conjecture that, for higher-spin gauge theories, the algebra of Hermitian differential operators, $$\textsc{T}=\frac12\left(\sum\limits_r\,(-i)^{r}\xi^{\mu_1\ldots\mu_{r}}(x)\,\partial_{\mu_1}\ldots
\partial_{\mu_{r}}\,+\,\mbox{Hermitian conjugate}\right)\,,$$ generalizes the algebra of infinitesimal diffeomorphisms for gravity. Another argument in favour of this conjecture may be presented in the ``gauge'' sector by looking at the metric-like formulation of higher-spins arising from the frame-like formulation of Vasiliev, at first order in the coupling constant \cite{Progress}.

\section{Conclusion}

The conclusion is that there are two complementary but distinct ways of
using rigid symmetries of the free theory in order to guess
the proper gauge symmetry principle of higher-spin gauge theories.

On the one hand, the infinite set of rigid symmetries of the free 
(or, maybe, even integrable) ``matter" sector, 
might be gauged by the introduction of a connection via a minimal coupling
prescription.
The idea of using a massive scalar field as free matter sector
and an infinite tower of massless symmetric tensor fields as
interacting gauge sector is in agreement with the isomorphism between
the off-shell higher-spin algebra and the space of reducibility parameters.
(If tensor fields are used as free ``matter'' sector, then
the symmetry algebra could be larger. Following the lines of the Vasiliev construction
in such case, the structure of the universal enveloping algebra
points towards a larger infinite tower of gauge
fields including mixed-symmetry tensors.)

On the other hand, in the free ``gauge" sector, rigid symmetries
linked to reducibility parameters may arise
from the linearization of the gauge symmetries of some
non-linear action. Thus the complete knowledge of the rigid symmetries
of free higher-spin gauge theories
would indicate what can be the linearized connection.

\section*{Acknowledgments}

I. Bakas, G. Barnich, N. Boulanger, T. Damour and J. Remmel
are thanked for very useful exchanges.
The author is grateful to the organizers for their invitation to
this enjoyable meeting and the opportunity to present his lecture.
The Institut des Hautes \'Etudes Scientifiques de Bures-sur-Yvette
is acknowledged for its hospitality.


\begin{thebibliography}{99}

\bibitem{Intros}
D.~Sorokin,
AIP Conf.\ Proc.\  {\bf 767} (2005) 172 [{\tt hep-th/0405069}];\\
N.~Bouatta, G.~Compere and A.~Sagnotti,
in the proceedings of the ``First Solvay Workshop on Higher-Spin
Gauge Theories" (Brussels, Belgium; May 2004) [{\tt hep-th/0409068}].

\bibitem{Ortin}
T.~Ortin, \textit{Gravity and strings} (Cambridge, 2004).

\bibitem{Noether}
F.~A.~Berends, G.~J.~H.~Burgers and H.~van Dam,
Nucl.\ Phys.\ B {\bf 260} (1985) 295.

\bibitem{Vasiliev}
M.~A.~Vasiliev,
Comptes Rendus Physique {\bf 5} (2004) 1101
[{\tt hep-th/0409260}];\\
X.~Bekaert, S.~Cnockaert, C.~Iazeolla and M.~A.~Vasiliev, 
in the proceedings of the ``First Solvay Workshop on
Higher-Spin Gauge Theories'' (Brussels, Belgium; May 2004) [{\tt
hep-th/0503128}].

\bibitem{Killingconstcurvv}
G. Thompson,
J. Math. Phys. {\bf 27} (1986) 2693;\\
R.~G.~McLenaghan, R.~Milson and R.~G.~Smirnov, C.\ R.\ Acad.\
Sci.\ Paris, Ser. {\bf I 339} (2004) 621.

\bibitem{BH}
G.~Barnich and M.~Henneaux,
Phys.\ Lett.\ B {\bf 311} (1993) 123 [{\tt hep-th/9304057}];\\
M.~Henneaux,
Contemp.\ Math.\  {\bf 219} (1998) 93 [{\tt hep-th/9712226}].

\bibitem{BBCL}
X.~Bekaert, N.~Boulanger, S.~Cnockaert and S.~Leclercq,
Fortsch.\ Phys.\  {\bf 54} (2006) 282 [{\tt hep-th/0602092}].

\bibitem{X}
X.~Bekaert, 
in the proceedings of the ``First Modave Summer School in
Mathematical Physics'' (Modave, Belgium; June 2005).

\bibitem{Littlewood}
D.E. Littlewood, \textit{The theory of group characters and matrix representations of groups} (Clarendon, 1958);\\
G.~R.~E.~Black, R.~C.~King and B.~G.~Wybourne,
J.\ Phys.\ A:\ Math.\ Gen.\ {\bf 16} (1983) 1555.

\bibitem{Mikhailov}
A.~Mikhailov,
``Notes on higher spin symmetries,''
{\tt hep-th/0201019}.

\bibitem{Zamolodchikov}
A.~B.~Zamolodchikov and A.~B.~Zamolodchikov,
Annals Phys.\  {\bf 120} (1979) 253.

\bibitem{Vasiliev:1995}
M.~A.~Vasiliev,
Int.\ J.\ Mod.\ Phys.\  D {\bf 5} (1996) 763
[{\tt hep-th/9611024}].

\bibitem{Eastwood}
R.~Geroch, J.\ Math.\ Phys.\ {\bf 11} (1970) 1955;\\
M.~G.~Eastwood, ``Higher symmetries of the Laplacian,'' {\tt
hep-th/0206233}.

\bibitem{Konstein}
S.~E.~Konstein, M.~A.~Vasiliev and V.~N.~Zaikin,
JHEP {\bf 0012} (2000) 018 [{\tt hep-th/0010239}].

\bibitem{Bakas}
I.~Bakas, B.~Khesin and E.~Kiritsis,
Commun.\ Math.\ Phys.\  {\bf 151} (1993) 233.

\bibitem{Dodd}
R.~K.~Dodd and R.~K.~Bullough,
Proc.\ Roy.\ Soc.\ Lond.\ A {\bf 352} (1977) 481.

\bibitem{Barnich}
G.~Barnich and F.~Brandt,
Nucl.\ Phys.\  B {\bf 633} (2002) 3
[{\tt hep-th/0111246}].

\bibitem{current}
D.~Anselmi,
Class.\ Quant.\ Grav.\  {\bf 17} (2000) 1383
[{\tt hep-th/9906167}];\\
M.~A.~Vasiliev, 
in M. Shifman ed., \textit{The many faces of the
superworld} (World Scientific, 2000) [{\tt hep-th/9910096}].

\bibitem{FS}
D.~Francia and A.~Sagnotti,
Class.\ Quant.\ Grav.\  {\bf 20} (2003) S473
[{\tt hep-th/0212185}];
J.\ Phys.\ Conf.\ Ser.\  {\bf 33} (2006) 57
[{\tt hep-th/0601199}].

\bibitem{FMS}
D.~Francia, J.~Mourad and A.~Sagnotti,
Nucl.\ Phys.\  B {\bf 773} (2007) 203
[{\tt hep-th/0701163}];\\
I.~L.~Buchbinder, A.~V.~Galajinsky and V.~A.~Krykhtin,
Nucl.\ Phys.\  B {\bf 779} (2007) 155
[{\tt hep-th/0702161}].

\bibitem{Progress}
X.~Bekaert, work in progress.


\end{thebibliography}
\end{document}